\documentclass[%
superscriptaddress,
preprint,
 amsmath,amssymb,
 aps,
]{revtex4-2}

\usepackage{graphicx}
\usepackage{dcolumn}
\usepackage{bm}
\usepackage{multirow}
\usepackage[version=3]{mhchem}

\usepackage{amsmath,esint}
\usepackage{verbatim}
\usepackage{spverbatim}
\RequirePackage[normalem]{ulem}
\RequirePackage{color}\definecolor{RED}{rgb}{1,0,0}\definecolor{BLUE}{rgb}{0,0,1}\definecolor{GREEN}{rgb}{0,1,0}

\begin{document}


\title{Band unfolding with a general transformation matrix: from code implementation to interpretation of photoemission spectra}

\author{Oleg Rubel}
\email[O.R. email: ]{rubelo@mcmaster.ca, ORCID: 0000-0001-5104-5602}
\affiliation{Department of Materials Science and Engineering, McMaster University, 1280 Main Street West, Hamilton, Ontario L8S 4L8, Canada}

\author{Jean-Baptiste Moussy}
\affiliation{Universit\'{e} Paris-Saclay, CEA, CNRS, SPEC, 91191, Gif-sur-Yvette, France}

\author{Paul Foulquier}
\affiliation{Universit\'{e} Paris-Saclay, CEA, CNRS, SPEC, 91191, Gif-sur-Yvette, France}
\affiliation {Universit\'{e} Paris-Saclay, CNRS, Laboratoire de Physique des Solides, 91405, Orsay, France.}

\author{V\'{e}ronique Brouet}
\email[V.B. email: ]{veronique.brouet@u-psud.fr}
\affiliation {Universit\'{e} Paris-Saclay, CNRS, Laboratoire de Physique des Solides, 91405, Orsay, France.}

\date{\today}

\begin{abstract}
Unfolding of a supercell band structure into a primitive Brillouin zone is important for understanding implications of structural distortions, disorder, defects, solid solutions on materials electronic structure. Necessity of the band unfolding is also recognised in interpretation of angle-resolved photoemission spectroscopy (ARPES) measurements. We describe an extension of the \texttt{fold2Bloch} package by implementing an arbitrary transformation matrix used to establish a relation between primitive cell and supercell. This development allows us to overcome limitations of supercells constructed exclusively by scaling of primitive cell lattice vectors. It becomes possible to transform between primitive and conventional cells as well as include rotations. The \texttt{fold2Bloch} is publicaly available from a GitHub repository as a FORTRAN code. It interfaces with the all-electron full-potential WIEN2k and the pseudopotential VASP density functional theory packages. The \texttt{fold2Bloch} is supplemented by additional pre- and post-processing utilities that aid in generating k points in the supercell (such that they later fall onto a desired path in the primitive Brillouin zone after unfolding) and plotting the unfolded band structure. We selected \ce{Sr2IrO4} as an illustrative example and, for the first time, present its properly unfolded band structure in direct comparison with ARPES measurements. In addition, critical importance of the band unfolding for interpretation of \ce{SrIrO3} ARPES data is illustrated and discussed as a perspective.
\end{abstract}

\maketitle


\section{\label{sec:Introduction}Introduction}

The band dispersion $E(k)$ obtained from electronic structure calculations provides information about Fermi surface of metals, direct/indirect transition in semiconductors, effective masses, etc. When periodicity is perturbed (e.g., solid solutions, defects, magnetic order) the Brillouin zone (BZ) shrinks and bands get folded. As a result, $E(k)$ becomes obscured and difficult to interpret.

Even when the periodicity is formally perturbed, it is still possible to recover an effective band structure in a BZ of the primitive cell using a k-spectral decomposition also known as `band unfolding'. There are numerous examples of band unfolding. The notable milestones include an electronic structure of aperiodic solids  \cite{Niizeki_JPCM_2_1990,Kosugi_JPSJ_86_2017,Mayo_JPCM_32_2020}, solid solutions described within a tight-binding approximation \cite{Dargam_PRB_56_1997,Boykin_PRB_71_2005,Boykin_JPCM_19_2007} or  pseudopotentials with a plane-wave basis set \cite{Wang_PRL_80_1998,Popescu_PRL_104_2010,Popescu_PRB_85_2012}, and Wannier functions derived from first-principle calculations to interpret angle-resolved photoemission spectroscopy (ARPES) experiments \cite{Ku_PRL_104_2010}. There are a number of public implementations of band unfolding in the density functional theory (DFT) community, for instance, \texttt{BandsUP} \cite{Medeiros_PRB_89_2014,Medeiros_PRB_91_2015}, \texttt{fold2Bloch} \cite{Rubel_PRB_90_2014}, \texttt{VASPKIT} \cite{Wang_CPC_267_2021}, \texttt{vaspwfc} \cite{Zheng_GitHub_PyVaspwfc_2022}.

The simplest scenario to unfold the band structure involves the supercell constructed from primitive real-space lattice vectors $\mathbf{a}_i$ ($i=1,2,3$) by scaling them with an integer $n_i \mathbf{a}_i$ ($n_i \in \mathbb{Z}^+$). However, it does not cover all possibilities. There are lattices constructed by a combination of rotation and scaling of the primitive structural unit. Examples include a conventional vs reduced unit cell \cite{Gruber_further-properties_2006,deWolff_red-basis_2006} or an octahedral rotation and tilting in perovskite structures \cite{Glazer_ACB_28_1972}. \citet{Popescu_PRB_85_2012} suggested a more general approach to unfolding that involves a transformation matrix. It was already implemented in \texttt{BandsUP}, \texttt{VASPKIT}, and \texttt{vaspwfc} but not in \texttt{fold2Bloch}.

\section{Method}\label{sec:Method}

\subsection{Lattice transformations}

Following the notations in Ref.~\citenum{Popescu_PRB_85_2012} we denote
by small (capital) symbols quantities referring to the primitive cell (supercell). Transformation of real-space primitive $\mathbf{a}_i$ to supercell $\mathbf{A}_i$ lattice vectors can be expressed as \cite{Arnold_transform-coord-sys_2006}
\begin{equation}\label{eq:a_p -> a_s}
    \mathbf{A}_{i} = \sum_{j=1}^3 P_{ji} \, \mathbf{a}_{j}  \quad (i=1,2,3)
\end{equation}
or in the matrix form as
\begin{equation}\label{eq:a_p -> a_s (matrix)}
    A = P^{\intercal} \, a,
\end{equation}
where the lattice vectors matrices $A$ and $a$ are constructed of rows being individual vectors and columns being their Cartesian components ($x,y,z$), e.g.,
\begin{equation}\label{eq:a_s (matrix)}
    A = 
    \begin{pmatrix}
        \mathbf{A}_{1} \\
        \mathbf{A}_{2} \\
        \mathbf{A}_{3}
    \end{pmatrix}
    \equiv
    \begin{bmatrix}
        A_{11} & A_{12} & A_{13} \\
        A_{21} & A_{22} & A_{23} \\
        A_{31} & A_{32} & A_{33}
    \end{bmatrix}.
\end{equation}
Here $P$ is a transformation matrix of the size $3 \times 3$ compatible with conventions recommended by the International tables for crystallography \cite{Arnold_transform-coord-sys_2006} (same as in VESTA structure visualization software \cite{Momma_JAC_44_2011} or Bilbao crystallographic server \cite{Aroyo_ACA_70_2014} but different from Ref.~\citenum{Popescu_PRB_85_2012}). The transformation matrix is obtained by solving the linear Eq.~\eqref{eq:a_p -> a_s (matrix)}, which yields
\begin{equation}\label{eq:P}
    P  = (A \, a^{-1})^{\intercal}.
\end{equation}
The reverse transformation of a supercell to the primitive cell is obtained using the inverse matrix $P^{-1}$
\begin{equation}\label{eq:a_s -> a_p (matrix)}
    a = (P^{-1})^{\intercal} \, A.
\end{equation}
Scaling of the cell volume as a result of the primitive cell to supercell transformation is given by \cite{Arnold_transform-coord-sys_2006}
\begin{equation}\label{eq:V_s/V_p}
    n_{\text{v}} = \det (P),
\end{equation}
which imposes two constrains: $P$ should be positive defined and $P_{ij} \in \mathbb{Z}$, from which it follows that $\det(P)\in \mathbb{Z}^+$.

Reciprocal lattice vectors are also transformed using $P$. Owing to the relation $B=(A^{-1})^{\intercal}$, reciprocal lattice vectors of a supercell $\mathbf{B}_i$ can be transformed to the reciprocal primitive vectors $\mathbf{b}_i$ as \cite{Arnold_transform-coord-sys_2006}
\begin{equation}\label{eq:b_s -> b_p (matrix)}
    b = P \, B
\end{equation}
and back as
\begin{equation}\label{eq:b_p -> b_s (matrix)}
    B = P^{-1} \, b.
\end{equation}
Note that the transposition of $P$ is not required for converting reciprocal lattice vectors in the matrix form contrary to the conversions proposed in Ref.~\citenum{Popescu_PRB_85_2012} (see Eqs.~(1) and (2) therein).

\subsection{Band unfolding}

We already outlined in Ref.~\citenum{Rubel_PRB_90_2014} an unfolding procedure used in the \texttt{fold2Bloch} implementation when a supercell is constructed by simple scaling of the primitive cell . Here we present an extended (more general) version.

DFT codes for solids internally operate with wave vectors in fractional coordinates. Here we will use tilde to denote primitive (supercell) fractional coordinates $\tilde{\mathbf{k}}$ ($\tilde{\mathbf{K}}$) where components of each vector span a range between 0 and 1. Cartesian components of the wave vector are obtained by a multiplication with the reciprocal lattice matrix
\begin{equation}\label{eq:k_p -> Cartesian}
    \mathbf{k} = \tilde{\mathbf{k}} \, b
\end{equation}
and similarly for the supercell $\mathbf{K}$.

\begin{figure}
	\includegraphics{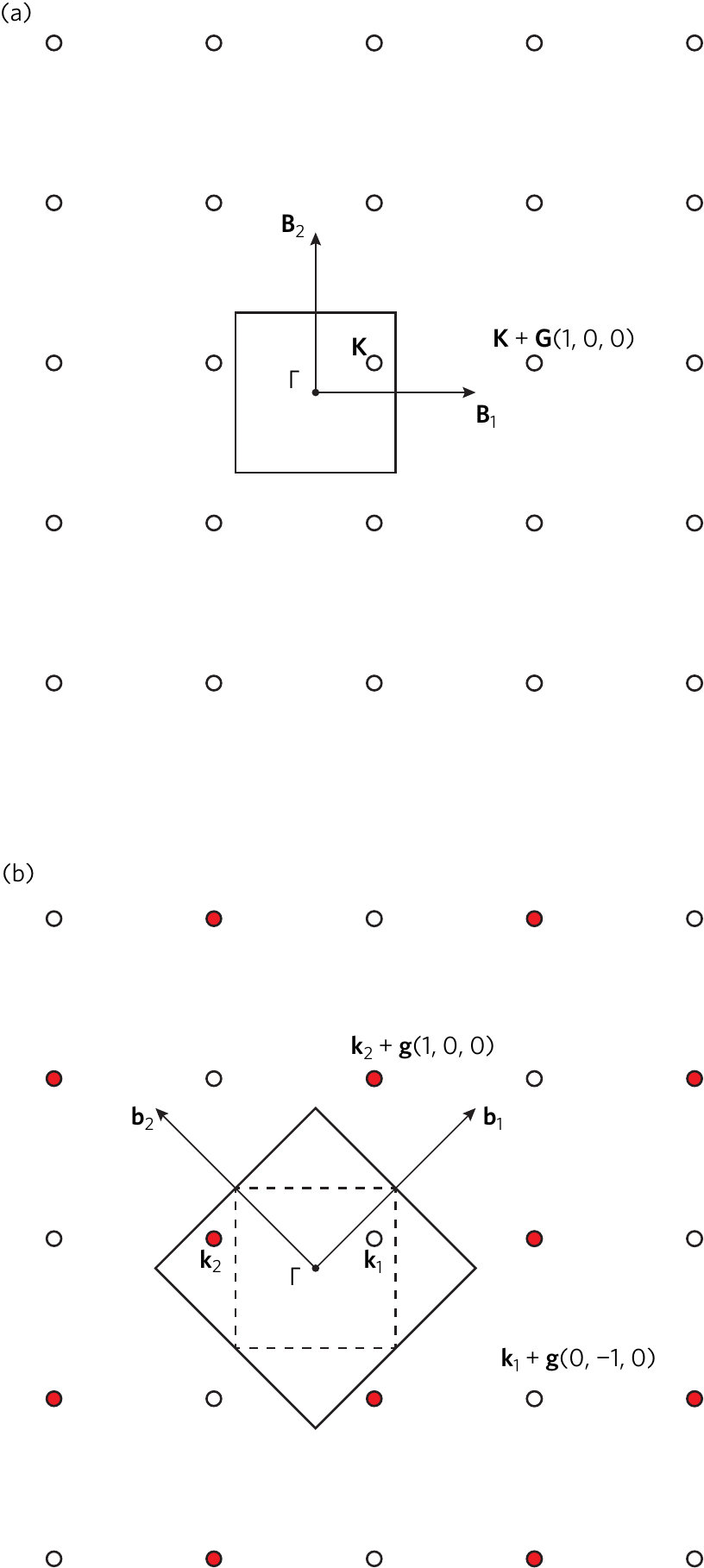}
	\caption{Unfolding in reciprocal space. (a) BZ of a supercell with an arbitrary $\mathbf{K}$ point. (b) BZ of a primitive cell obtained by the two-dimensional transformation matrix of $P=[1 , 1 ; \bar{1}, 1]$. The cell volume changes twice ($n_{\text{v}}=2$), thus the $\mathbf{K}$ point transforms into two new point $\mathbf{k}_1$ and $\mathbf{k}_2$ that form two groups of plane wave coefficients (open and red markers). The dashed line shows the supercell BZ inside of the primitive BZ.}\label{fig-unfolding-rotation}
\end{figure}

Transformation of reciprocal-space supercell $\tilde{\mathbf{K}}$ to primitive $\tilde{\mathbf{k}}$ wave vectors (fractional coordinates) is given by \cite{Arnold_transform-coord-sys_2006}
\begin{equation}\label{eq:K_s -> k_p}
    \tilde{\mathbf{k}} =  [\tilde{\mathbf{K}} + (m_1,m_2,m_3)] \, P^{-1} \mod 1 .
\end{equation}
With all possible combinations of $m_i \in \mathbb{Z}$, the number of unique $\mathbf{k}$ points generated within the first BZ of the primitive cell (Fig.~\ref{fig-unfolding-rotation}) is given by the volume scale (Eq.~\eqref{eq:V_s/V_p}). The new (unfolded) k points in Fig.~\ref{fig-unfolding-rotation}b form two subsets $\mathbf{k}_1$ (open markers) and $\mathbf{k}_2$ (red) of the original grid $\mathbf{K} + \mathbf{G}$ in Fig.~\ref{fig-unfolding-rotation}a. The two subsets were created since the volume change is $n_{\text{v}}=2$ in the example shown. More generally, the subset property is expressed as
\begin{equation}\label{eq:k as a subset of K+G}
    \mathbf{k}_l + \mathbf{g} \subset \mathbf{K} + \mathbf{G} \quad (l=1, \ldots, n_{\text{v}}),
\end{equation}
with $\mathbf{G}(m_1,m_2,m_3) = m_1 \mathbf{B}_1 + m_2 \mathbf{B}_2 + m_3 \mathbf{B}_3$ and $\mathbf{g}(m_1,m_2,m_3) = m_1 \mathbf{b}_1 + m_2 \mathbf{b}_2 + m_3 \mathbf{b}_3$ being commensurate vectors of the plane wave expansion.

The wave function in WIEN2k is split into two regions: the atomic spheres and the interstitial region \cite{Blaha_WIEN2k_2018}. A plane wave basis set is used in the interstitial region 
\begin{equation}\label{eq:PW wave function}
    \Psi_{\sigma, n,\mathbf{K}}^{(\text{int})}(\mathbf{r}) = 
    \sum_{\mathbf{G}} C_{\sigma, n,\mathbf{K}}(\mathbf{G}) \,
    e^{i(\mathbf{K} + \mathbf{G}) \cdot \mathbf{r}},
\end{equation}
where $C$ are plane wave coefficients for a specific electronic state with the wave vector $\mathbf{K}$, spin channel $\sigma$, band index $n$.

Spectral weight of the new unfolded k point is evaluated from the subset of plain wave coefficients \cite{Popescu_PRB_85_2012}
\begin{equation}\label{eq:w(s,n,k)}
    w_{\sigma, n} (\mathbf{k}_l) = \sum_{\mathbf{g}} |C_{\sigma, n,\mathbf{K}}(\mathbf{k}_l + \mathbf{g})|^2 \quad (l=1, \ldots, n_{\text{v}}).
\end{equation}
Weights are normalized such that
\begin{equation}\label{eq:w(s,n,k) norm}
    \sum_{l=1}^{n_{\text{v}}} w_{\sigma, n} (\mathbf{k}_l) = 1
\end{equation}
In the case of a spinor wave function, weights of spin up and down components are mixed
\begin{equation}\label{eq:w(n,k) spinor}
    w_{n} (\mathbf{k}) = \alpha^2 w_{\uparrow ,n} (\mathbf{k})  + \beta^2 w_{\downarrow ,n} (\mathbf{k}),
\end{equation}
where $\alpha$ and $\beta$ are components of the spinor wave function ($\alpha^2 + \beta^2 = 1$). This result is similar to decomposition of partial spectral weights proposed by \citet{Medeiros_PRB_91_2015}, yet it is additionally augmented by the relative contribution of each spin channel to the wave function.

\subsection{Electronic structure calculations}

All electronic structure calculations were performed with WIEN2k DFT package \cite{Blaha_WIEN2k_2018,Blaha_JCP_152_2020}. Relevant parameters are listed in Table~\ref{tab-param}. We used the \citet*{Perdew_PRL_77_1996} (PBE) exchange-correlation functional in combination with the onsite Hubbard correction $U$ \cite{Anisimov_PRB_48_1993} for Ir-d electrons in the case of \ce{Sr2IrO4}. The spin-orbit coupling was included in all calculations. The spin polarization was enabled only in the case of \ce{Sr2IrO4}, where we used a collinear antiferromagnetic ordering as in Ref.~\citenum{Ye_PRB_87_2013} (Fig.~\ref{fig-Sr2IrO4-structure}d) initialized with the magnetic moment of $\pm 1$ Bohr magneton ($\mu_{\text{B}}$) per Ir site.

\subsection{Implementation and execution workflow}

\textbf{WIEN2k}: First we initialize the calculation and complete the self-consistent field (SCF) cycle using the \texttt{case.struct} structure file as an input to obtain a self-consistent potential and a charge density. Once the calculation converges all necessary files are saves in the \texttt{SOC} folder. A detailed workflow can be found in the supporting information (SI) section.

\textbf{Utils/fold.m}: Next we generate a list of \textit{folded} $\tilde{\mathbf{K}}$ points within the supercell BZ using the Octave/MATLAB script that takes a desired $\tilde{\mathbf{k}}$ point path in the primitive BZ, the number of intermediate points for each section of the path, as well as the transformation matrix $P$ as inputs. \\
\texttt{\$ octave fold.m}\\
The folding is achieved by the following matrix product \cite{Arnold_transform-coord-sys_2006}
\begin{equation}\label{eq:K_p -> k_s}
    \tilde{\mathbf{K}} =  \tilde{\mathbf{k}} \, P \mod 1 .
\end{equation}
The generated unique $\tilde{\mathbf{K}}$ points are stored in the \texttt{case.klist\_band} file in a WIEN2k native format as three integer numbers per k point with a common integer divisor. It is important to note conventions used within WIEN2k to interpret k~point coordinates in the \texttt{case.klist\_band} file. The BZ of a \textit{conventional} lattice is implied for F, B, CXY, CXZ, and CXZ orthorhombic lattices. The BZ of a \textit{primitive} lattice is used for other lattice types (P, H, R, CXZ monoclinic). Those peculiarities affect the selection of supercell lattice vectors $A$ in Eq.~\eqref{eq:a_s (matrix)} and the construction of $P$ matrix using Eq.~\eqref{eq:P}. In practice, we expect that majority of users will have P-type supercells due the symmetry broken by disorder/defects.

\textbf{WIEN2k}: Now we generate eigenvalues and eigenvectors (wave functions or vector files) for the list of k points in the \texttt{case.klist\_band} file. We use files saved in the \texttt{SOC} folder after the previous SCF step.\\
\texttt{\$ x lapw1 -band -up [-p]}\\
\texttt{\$ x lapw1 -band -dn [-p]}\\
\texttt{\$ x lapwso -up [-orb] [-p]}\\
The \texttt{-orb} switch is activated for the DFT+$U$ calculation. Here we produce files that are essential for unfolding: \texttt{case.vectorso[up/dn]} (wave functions and energy eigenvalues) and \texttt{case.normso[up/dn]} (spinor components $\alpha^2$ and $\beta^2$).

\textbf{fold2Bloch}: fold2Bloch is a FORTRAN code that can be compiled with either Intel or GNU FORTRAN compilers. It takes wave functions, spinor components, and the transformation matrix $P$ as input arguments\\
\texttt{\$ fold2Bloch -so case.vectorsoup[\_1] case.vectorsodn[\_1]}\\
\texttt{... case.normsoup[\_1] case.normsodn[\_1]}\\
\texttt{... "'P11 P12 P13:P21 P22 P23:P31 P32 P33'"}\\
and generates a \texttt{case.f2b} file. If WIEN2k calculations run in a k-parallel mode (\texttt{[-p]} option), output vector and norm files will be marked with \texttt{\_XX} for each parallel stream. These files can be processed individually, and the output can be concatenated into one \texttt{case.f2b} file. The sample listing of the output file is given below
\begin{verbatim}
     k_1        k_2        k_3       E (Ry)       w
...
   0.000000   0.000000   0.000000   0.393800   0.001619
   0.000000   0.000000   0.250000   0.393800   0.000001
   0.000000   0.000000   0.500000   0.393800   0.487296
   0.000000   0.000000   0.750000   0.393800   0.000001
   0.500000   0.500000   0.000000   0.393800   0.000000
   0.500000   0.500000   0.250000   0.393800   0.255542
   0.500000   0.500000   0.500000   0.393800   0.000000
   0.500000   0.500000   0.750000   0.393800   0.255542
   0.000000   0.000000   0.000000   0.394518   0.374261
   0.000000   0.000000   0.250000   0.394518   0.000001
   0.000000   0.000000   0.500000   0.394518   0.001956
   0.000000   0.000000   0.750000   0.394518   0.000001
   0.500000   0.500000   0.000000   0.394518   0.000000
   0.500000   0.500000   0.250000   0.394518   0.311889
   0.500000   0.500000   0.500000   0.394518   0.000000
   0.500000   0.500000   0.750000   0.394518   0.311893
...
\end{verbatim}
Here one can see results of band unfolding with the transformation matrix $P=[1,\bar{1},\bar{2};1,1,\bar{2};0,0,4]$ and the volume scale of $n_{\text{v}}=8$. Two groups of eigenvalues are shown each unfolded into eight new k points (fractional coordinates) in the primitive BZ. Even though both eigenvalues belong to $\Gamma$ point in the supercell BZ, after unfolding the first does not have any notable $\Gamma$ character (only ca.~0.16\%), while the second eigenvalue retains about 37\% of its $\Gamma$ character.

\textbf{Utils/ubs\_bmp.m}: This Octave script is used to prepare a binary file \texttt{case.f2b.bin} for band structure plotting. The inputs are the \texttt{case.f2b} file, the desired $\tilde{\mathbf{k}}$ path in the primitive BZ (it has to match that used as input in \texttt{fold.m}), reciprocal lattice vectors of the supercell (can be read from the \texttt{case.outputkgen} file in a column-wise manner and later transposed within the script), the Fermi energy from \texttt{case.scf} file, smearing for a Gaussian function in energy and k space.\\
\texttt{\$ octave ubs\_bmp.m}\\
Sensible values for the energy and k space smearing are about 1/50 of the energy range and the k path length selected for the band structure plot. At the end of execution, \texttt{XTICKS} and \texttt{KLABEL} vectors are printed. They need to be noted and used in the next stage.

\textbf{Utils/f2b-band-structure.plt}: Finally, the unfolded band structure is plotted using Gnuplot with the input binary file \texttt{case.f2b.bin}.\\
\texttt{\$ gnuplot f2b-band-structure.plt}\\
The Gnuplot script incorporates data from \texttt{XTICKS} and \texttt{KLABEL} vectors used for labeling the high-symmetry points along the path and generates a publication-quality plot (\texttt{case.eps}).

\subsection{Experimental details}

The experimental structure of \ce{Sr2IrO4} was measured with ARPES on the Cassiop{\'e}e beamline of the SOLEIL synchrotron, with a SCIENTA R-4000 analyzer and an overall resolution better than 15~meV. The temperature was 20~K, the photon energy was set to 100~eV and linear polarization in the plane containing $\Gamma$M was used.  The samples were prepared using a self-flux method, as reported before \cite{Brouet_PRB_92_2015,Louat_PRB_100_2019}. 

We have grown \ce{SrIrO3} thin films on \ce{SrTiO3} (001) substrate using pulsed laser deposition. A frequency-tripled Nd:YAG laser ($\lambda=355$~nm, $f=2.5$~Hz, pulse duration 15~ns) was focused on a polycrystalline \ce{SrIrO3} target made by solid state synthesis. The substrate surface was prepared following the process described in Ref.~\citenum{Connell_APL_101_2012} to obtain a uniform \ce{TiO2} surface termination. The deposition was performed with the substrate heated at $T=600~^{\circ}$C and with oxygen partial pressure $P=2.5 \times 10^{-1}$~mbar and monitored by in-situ reflection high-energy electron diffraction. Then, the thin films were cooled down to room temperature with the same oxygen partial pressure to compensate for any oxygen vacancy. Finally, we fully structurally and physically characterized our thin films by X-ray diffraction and reflectivity using a Br{\"u}ker D8 advance diffractometer and by surface diffraction on SIXS beamline at SOLEIL synchrotron, atomic force microscopy and electronic transport, concluding to similar characteristics and properties to previous reports in the literature \cite{Bhat_EPL_122_2018,GutierrezLlorente_AM_6_2018,Zhang_PRB_91_2015}.
 
\section{Results and Discussion}\label{sec:Results and Discussion}

\subsection{\ce{Sr2IrO4}: Theoretical calculations}

\begin{figure}
	\includegraphics{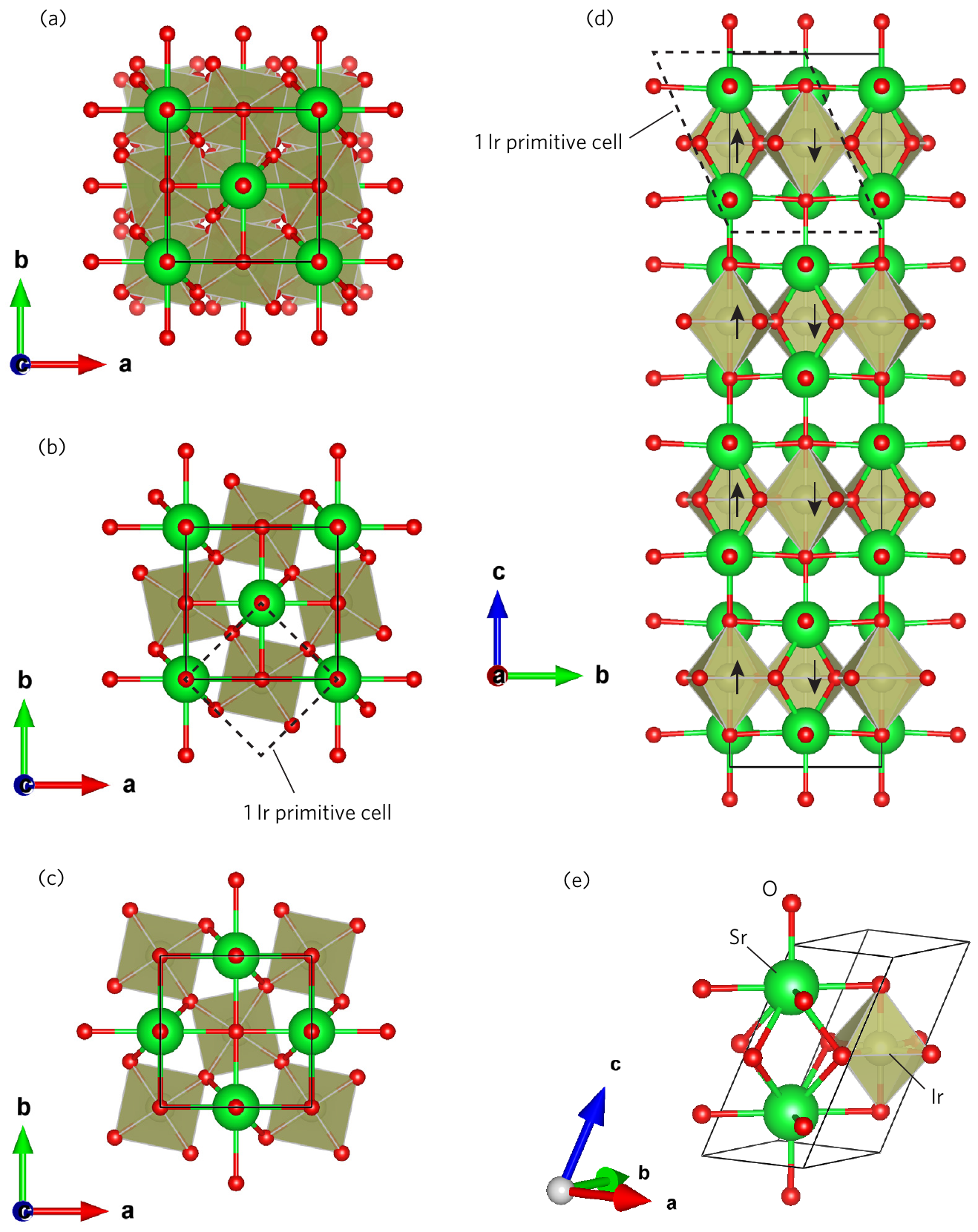}
	\caption{Structure of \ce{Sr2IrO4} (space group 88, I4$_1$/a (non-magnetic) or 13, P2/c (magnetic)): (a) top view, (b) top view with the first Ir layer only, (c) top view with the second Ir layer only, (d) side view with arrows showing the magnetic ordering, and (e) primitive 1-Ir unit cell (space group 139, I4/mmm) obtained from the supercell without octehedral tilting using the transformation matrix of $P^{-1}=(1/2,1/2,1/2;\overline{1/2},1/2,0;0,0,1/4)$. The octahedral tilting is present on panels (a)--(d).}\label{fig-Sr2IrO4-structure}
\end{figure}

The structure of \ce{Sr2IrO4} was imported from Springer Materials \cite{Ye_PRB_92_2015} (dataset ID~sd\_1945591). The original structure is BCT and includes a tilting of \ce{IrO6} octahedra  (Fig.~\ref{fig-Sr2IrO4-structure}a-d). The magnetic ordering further reduces the symmetry to a tetragonal cell with 8-Ir atoms. Its lattice vectors matrix ({\AA}) is
\begin{equation}
    A = 
    \begin{bmatrix}
        5.485 & 0 & 0 \\
        0 & 5.485 & 0 \\
        0 & 0 & 25.775
    \end{bmatrix}.
\end{equation}
For verification purposes we also created an idealized structure by eliminating the octahedral tilting and magnetic ordering. Its elementary structural unit (a 1-Ir primitive BCT unit cell) is shown in Fig.~\ref{fig-Sr2IrO4-structure}e. The corresponding matrix of primitive lattice vectors is
\begin{equation}
    a = 
    \begin{bmatrix}
        A_1/2 & -A_2/2 & 0 \\
        A_1/2 & A_2/2 & 0 \\
        A_1/2 & 0 & A_3/4
    \end{bmatrix}.
\end{equation}
Following Eq.~\eqref{eq:P} we obtain the transformation matrix
\begin{equation}
    P = 
    \begin{bmatrix}
        1 & -1 & -2 \\
        1 & 1 & -2 \\
        0 & 0 & 4
    \end{bmatrix},
\end{equation}
which establishes the relation between 1-Ir primitive cell and 8-Ir supercell (without distortions). The structure files (in a WIEN2k native format) can be found in the SI section and visualized with XcrySDen \cite{Kokalj_CMS_28_2003} or VESTA \cite{Momma_JAC_44_2011}.

\begin{figure}
	\includegraphics{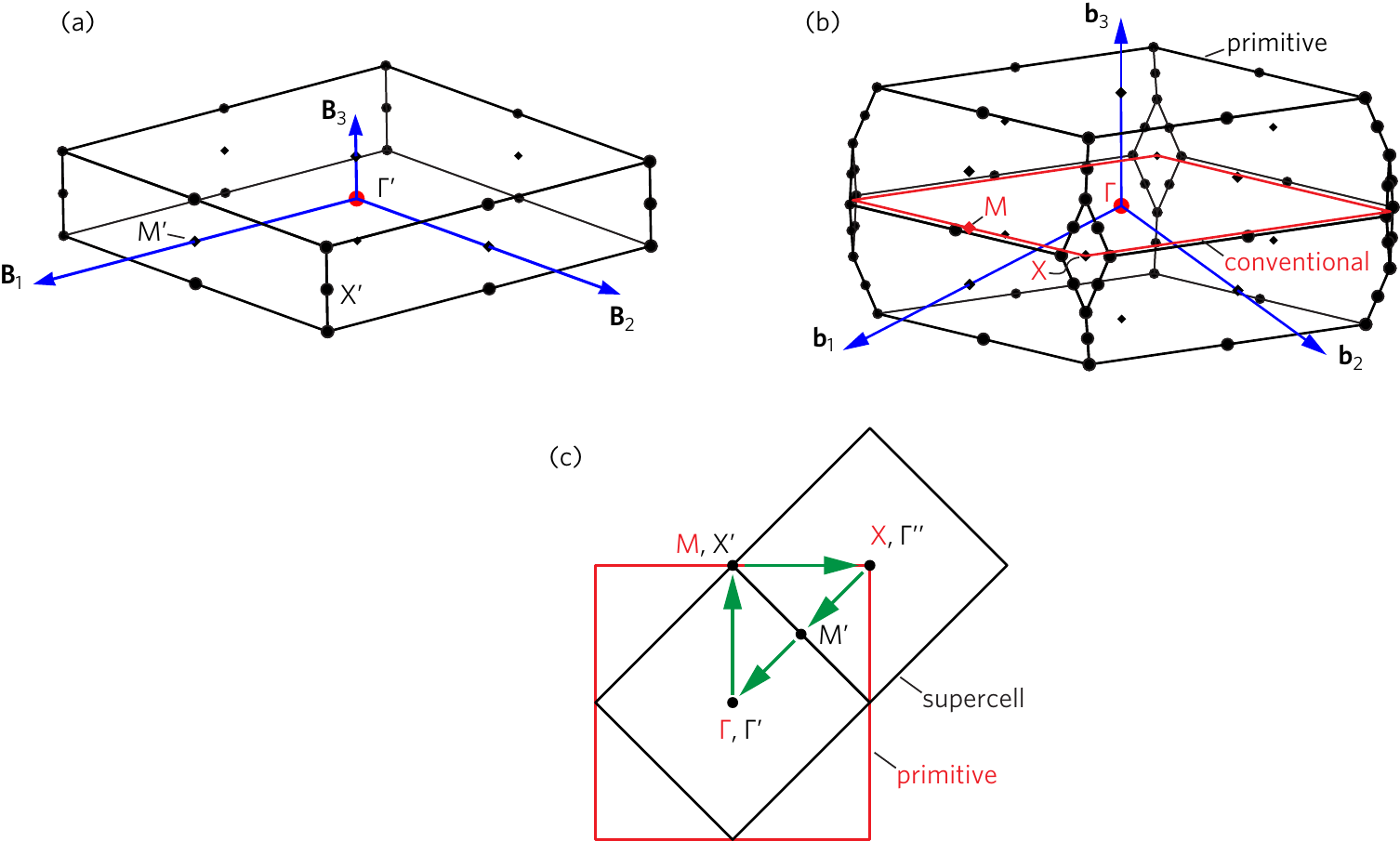}
	\caption{(a) BZ of \ce{Sr2IrO4} tetragonal cell (space group 88). (b) Primitive BZ of a body-centred tetragonal lattice (black) and the $(k_x, k_y, 0)$ plane of a conventional BZ (red). (c) Overlay of the primitive and supercell BZ (top view). The k path of interest within the $(k_x, k_y, 0)$ plane is shown by green arrows.}\label{fig-BZ-Sr2IrO4}
\end{figure}

The BZ of 8-Ir and 1-Ir cell is shown in Fig.~\ref{fig-BZ-Sr2IrO4}a,b, respectively. For plotting band structures we selected the $\Gamma\!-\!\text{M}\!-\!\text{X}\!-\!\Gamma$ path. Since we made a non-standard choice for BCT primitive lattice vectors (see Table~9.1.7.2, $tI$ Bravais lattice in the International tables for crystallography \cite{Burzlaff_bravais_lattices_2006}), special care should be taken to map $k$ point coordinates from the conventional to the primitive cell (Fig.~\ref{fig-BZ-Sr2IrO4}b) to ensure that the coordinates are compatible with the chosen definition for lattice vectors (Table~\ref{tab-BCT-k-conversion}).

\begin{figure}
    \includegraphics{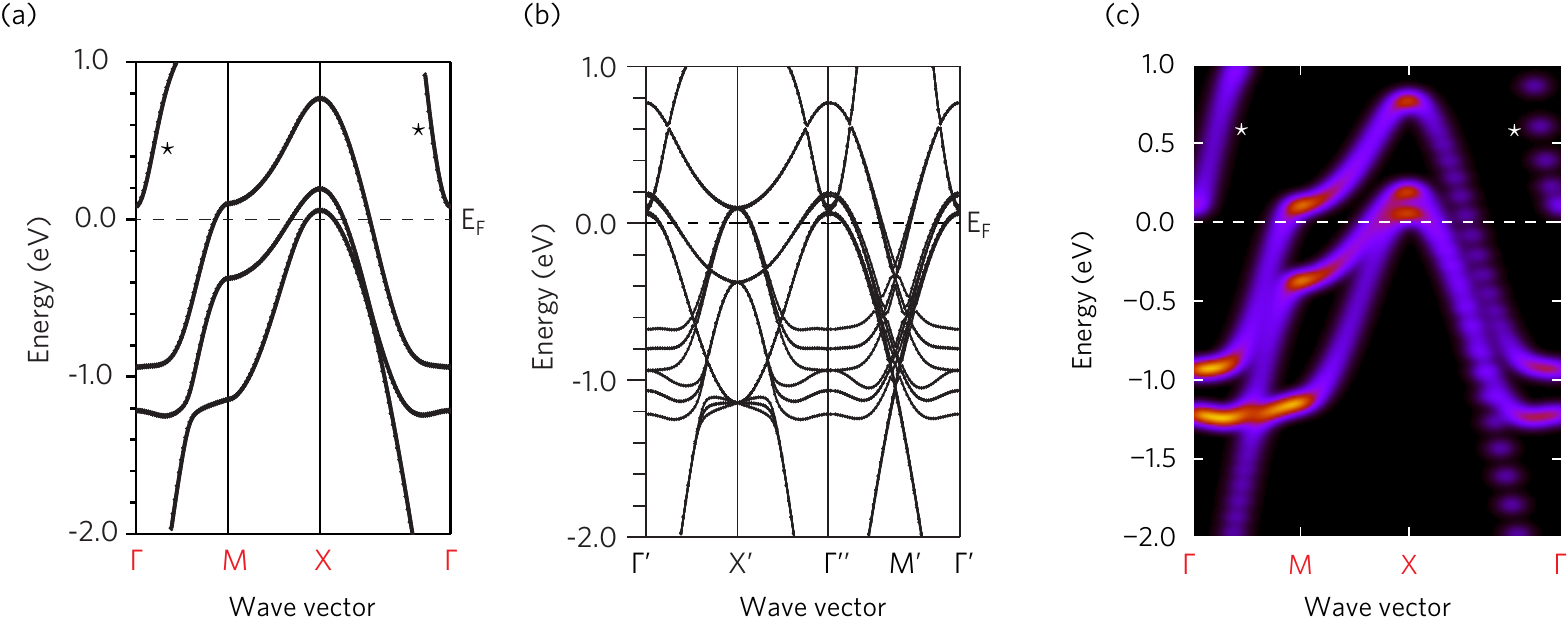}
    \caption{Band structure of \ce{Sr2IrO4} at PBE+SOC level of theory without octehedral tilting: (a) body-centred tetragonal primitive cell (space group 139, 1-Ir atom), (b) tetragonal supercell (space group 88, 8-Ir atoms), (c)  supercell unfolded into the primitive cell using $P=[1,\bar{1},\bar{2};1,1,\bar{2};0,0,4]$. The asterisk (*) marks $d_{x^2-y^2}$ states. Energies are plotted relative to the Fermi energy. Red (black) labels for high-symmetry points in the reciprocal space refer to the primitive cell (supercell) BZ.}\label{fig-bandstruct-Sr2IrO4-undistorted}
\end{figure}

For verification purposes we first need to calculate the band structure \textit{without} tilting of \ce{IrO6} octehedra. This way we can establish a direct comparison between the 8-Ir cell  and the primitive 1-Ir BCT cell. It is convenient to do this comparison at the PBE level of theory, since it leads to a non-magnetic solution with a more simple band structure (Fig.~\ref{fig-bandstruct-Sr2IrO4-undistorted}a). Bands near the Fermi energy are due to Ir-d electrons: the dispersive bands starting off $\Gamma$ near $E_{\text{F}}$ correspond to the $d_{x^2-y^2}$ orbital; the remaining bands are due to $t_{2g}$ states ($d_{xy}$, $d_{yz}$, and $d_{xz}$ orbitals); the $d_{z^2}$ orbital belongs to $e_g$ states located at higher energies outside the energy range of interest.

Figure~\ref{fig-bandstruct-Sr2IrO4-undistorted}b shows the band structure of the 8-Ir supercell without octahedral tilting. This band structure is obscured by the zone folding. Figure~\ref{fig-BZ-Sr2IrO4}c can help to rationalize zone folding at high-symmetry $k$ points (prime indices are for the supercell). Eigenvalues at both $\Gamma$ and X points are folded into $\Gamma '$ of the supercell. Eigenvalues at M point fall into $\mathrm{X}'$. The Dirac-like band crossings along the $\Gamma'\!-\!\mathrm{X}'$ segment are folding artefacts not observed in the primitive cell. The unfolded and primitive band structures are identical (Fig.~\ref{fig-bandstruct-Sr2IrO4-undistorted}a,c) that gives confidence in our approach and its implementation.

\begin{figure}
    \includegraphics{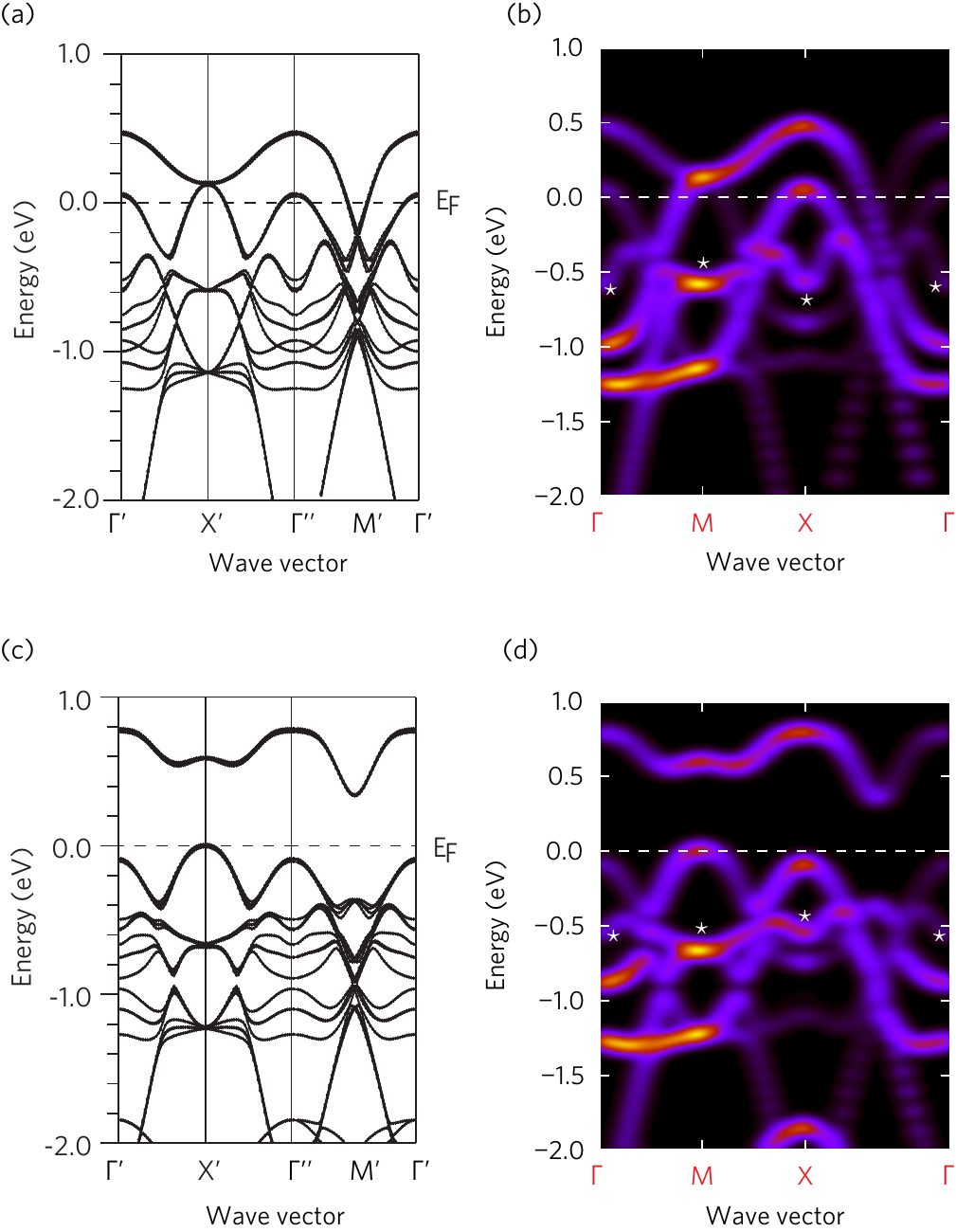}
    \caption{Unfolded band structure of \ce{Sr2IrO4} with octehedral tilting: (a,b) PBE+SOC folded and unfolded, (c,d) PBE+SOC with onsite $U_{\text{eff}}=3$~eV for Ir-d. Energies are plotted relative to the Fermi energy. The asterisk (*) marks $d_{xy}$ states.}\label{fig-bandstruct-Sr2IrO4-distorted}
\end{figure}

The realistic structure of \ce{Sr2IrO4} includes tilting of \ce{IrO6} octahedra. Those distortions cause perturbations in the band structure, which can be assessed thanks to the new functionality of \texttt{fold2Bloch}. The unfolded band structure of 8-Ir cell \textit{with} octahedral tiltings (Fig.~\ref{fig-bandstruct-Sr2IrO4-distorted}b) can now be compared to Fig.~\ref{fig-bandstruct-Sr2IrO4-undistorted}c (both calculated at the PBE level of theory). The most notable change is that the rotation allows hybridization between $d_{x^2-y^2}$ and $d_{xy}$, as is also the case in \ce{Sr2RhO4} \cite{Kim_PRL_97_2006}. Therefore, new $d_{x^2-y^2}$ states move to higher energies (not visible in Fig.~\ref{fig-bandstruct-Sr2IrO4-distorted}) and  $d_{xy}$ states are now pushed below the Fermi energy resulting in a new bright `spot' at X below the Fermi energy (Fig.~\ref{fig-bandstruct-Sr2IrO4-distorted}b), which is unfolded from $\Gamma''$ leaving a weak replica at $\Gamma$. Interestingly, states at M point are immune to the distortions.

To account for correlation effects we added the Hubbard $U_{\text{eff}}=3$~eV correction for Ir-d states. The magnitude of $U_{\text{eff}}$ was chosen to reproduce the experimental band gap of 0.5~eV \cite{Moon_PRL_101_2008,Battisti_NP_13_2016}. Unlike PBE, PBE+$U$ favours a magnetic solution with the moment of $\mu(\ce{Ir}) \approx 0.24 \mu_{\text{B}}$ per Ir site (the ordering is shown in Fig.~\ref{fig-Sr2IrO4-structure}d with the $[001]$ spin quantization direction for simplicity). Experimental $\mu(\ce{Ir})$ moments are 0.21~\cite{Ye_PRB_87_2013}, 0.29~\cite{Lovesey_JPCM_24_2012}, and 0.37~\cite{Dhital_PRB_87_2013}. A gap opens up between the on-site spin up and down states, which is particularly clear at M and at the middle of $\Gamma$X (Fig.~\ref{fig-bandstruct-Sr2IrO4-distorted}c-d).

\subsection{\ce{Sr2IrO4}: Comparison with experiment}

\begin{figure}
    \includegraphics{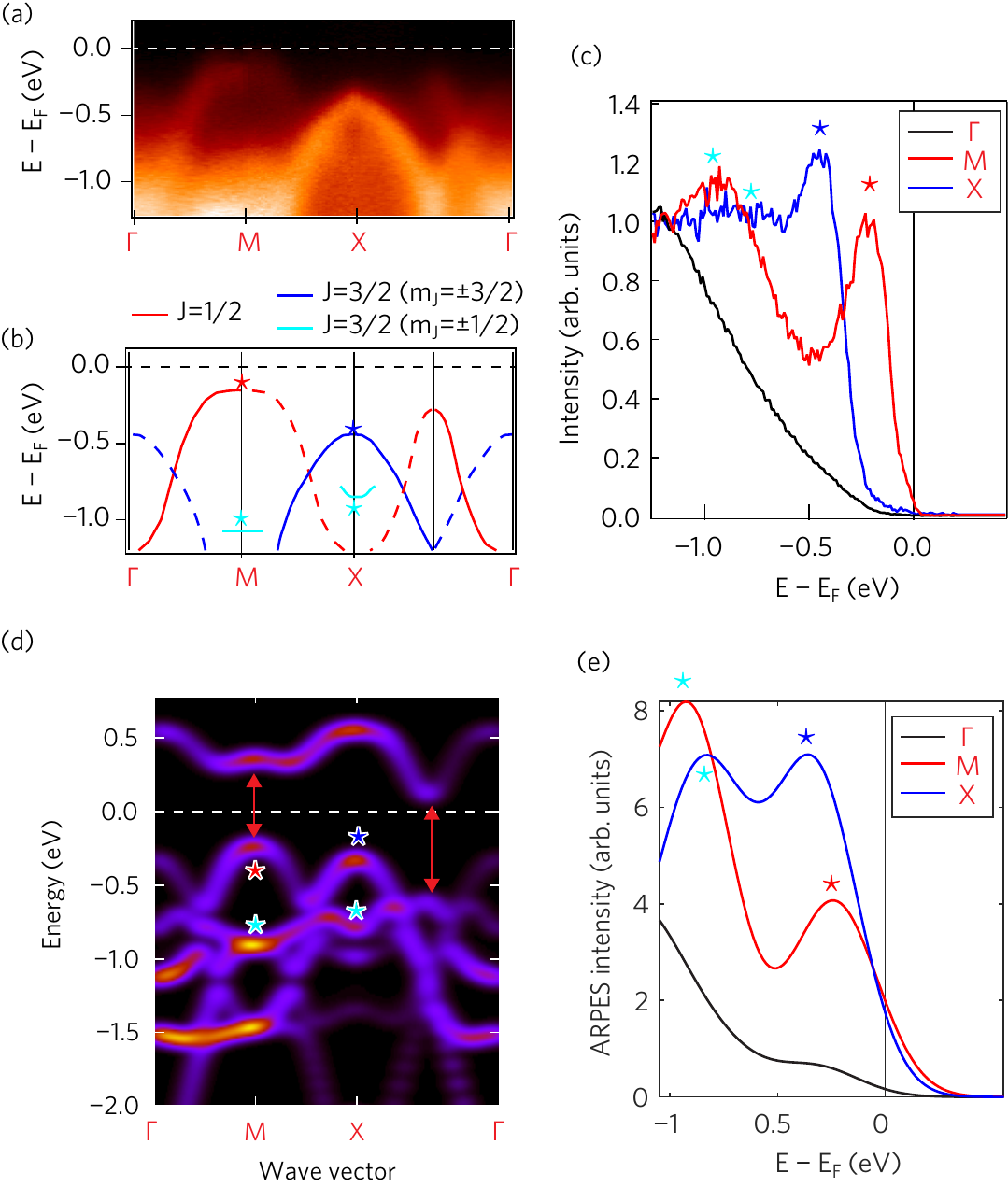}
    \caption{(a) Energy-momentum plot of the ARPES intensity as a function of the path \mbox{$\Gamma$-M-X-$\Gamma$}. (b) Sketch of the dispersion of the main bands visible in ARPES. The lines are guides to the eyes extracted from the data, the colors are given by comparison to the calculation. (c) Energy distribution curves at X (blue), $\Gamma$ (black) and M (red). (d) Comparison with the unfolded calculation along the same path.  The red arrow mark the opening of the magnetic gap. (e) Calculated ARPES spectra at X, $\Gamma$, and M (see text for details). Position of the Fermi energy in calculations was adjusted to match the experimental energy distance between the Fermi energy and the $J=1/2$ band maximum at M (red star) on the panel (c).} \label{fig-ARPES_Sr2IrO4}
\end{figure}

Figure~\ref{fig-ARPES_Sr2IrO4}(a) shows the ARPES spectral intensity along the k-path $\Gamma$-M-X-$\Gamma$, which was extracted from our measurement using a photon energy of 100~eV and a linear polarization along $\Gamma$M \cite{Brouet_PRB_92_2015}. The different bands observed in this plot are sketched in Fig.~\ref{fig-ARPES_Sr2IrO4}(b) by red and blue guides to the eyes, the colour corresponds to their dominant orbital character, characterized by the effective value of the total electronic angular momentum $J$, either $J=1/2$ ($m_J=\pm 1/2$) or $J=3/2$ ($m_J=\pm 1/2, \pm 3/2$). These data are similar to several previous reports  \cite{Kim_PRL_101_2008,Martins_PRM_2_2018,Louat_PRB_100_2019}, where more details can also be found. 

The most intense band is the $J=3/2$ band at X (blue star).  Although it should also be present at $\Gamma$, which is equivalent in the supercell (see Fig.~\ref{fig-BZ-Sr2IrO4}(c)), it can hardly be distinguished there. This is perfectly captured by the unfolded calculation of Fig.~\ref{fig-ARPES_Sr2IrO4}(d), which features much lower intensity for this band at $\Gamma$ compared to X. A comparison of the measured and calculated intensity along $\Gamma$ and X is also shown in Fig.~\ref{fig-ARPES_Sr2IrO4}(c) and (e), respectively. ARPES spectra in Fig.~\ref{fig-ARPES_Sr2IrO4}(e) were calculated based on the discrete spectral weights $w_n(\mathbf{k})$ defined by Eq.~\eqref{eq:w(n,k) spinor} and energies $E_n(\mathbf{k})$ of the unfolded band structure at specific $k$ points. A Gaussian broadening of the width $\sigma=0.2$~eV was applied. However, it is still a very crude approximation as other relevant details were omitted (matrix elements for initial- and final-state crystal wave functions, finite-lifetime effects, surface discontinuity, multiple scattering \cite{Damascelli_RMP_75_2003}). These matrix elements will further modulate the measured intensity, but the qualitative difference is well captured by the calculation. However, the relative position of the $J=3/2$ band at X and $J=1/2$ band at M is quite different in the two cases. In experiment, those bands are ca.~0.22~eV apart, while in calculations the energy difference between the valence band maxima at M and X is ca.~0.11~eV (compare red and blue stars on Fig.~\ref{fig-ARPES_Sr2IrO4} panels (c) and (e)). This discrepancy is due to an underestimation of the effective spin-orbit coupling already noticed and discussed in the literature \cite{Liu_PRL_101_2008,Zhou_PRX_7_2017}.

The $J=1/2$ band is the one where the magnetic gap opens \cite{Martins_PRM_2_2018} (Fig.~\ref{fig-ARPES_Sr2IrO4}(d), red arrows). This band is clearly visible along $\Gamma$M, but much weaker along MX, two paths expected to be equivalent in the supercell. Again, this fits with the theoretical calculation. Similarly, the $J=1/2$ band drastically loses weight on the second half of $\Gamma$X, both in the experiment and in calculation. Note that, as the relative positions of $J=1/2$ and $J=3/2$ are not correctly captured, the break in the dispersion due to hybridization between $J=1/2$ and $J=3/2$ where they cross is also shifted. 

The other bands are more difficult to isolate in ARPES, either because they become too broad at high binding energies or because they have low intensity in these experimental conditions. However, it is clear that another band is present at $-0.8$~eV at M and a trace of a second band at X can be seen. They are marked by cyan stars and correspond well to the other $J=3/2$ band ($m_J=\pm1/2$), which is dominated by $d_{xy}$ weight, and therefore has a lower cross section in ARPES \cite{Louat_PRB_100_2019}. 

\subsection{\ce{SrIrO3}: Perspective}

We illustrated the unfolding process with the case of \ce{Sr2IrO4}, where the connection to a primitive unit cell, without rotation of the oxygen octahedra, is relatively easy to anticipate. However, in more complicated cases, it can become totally impossible to understand the band structure and compare it to ARPES data, without the help of the unfolding calculation. 

\begin{figure}
    \includegraphics{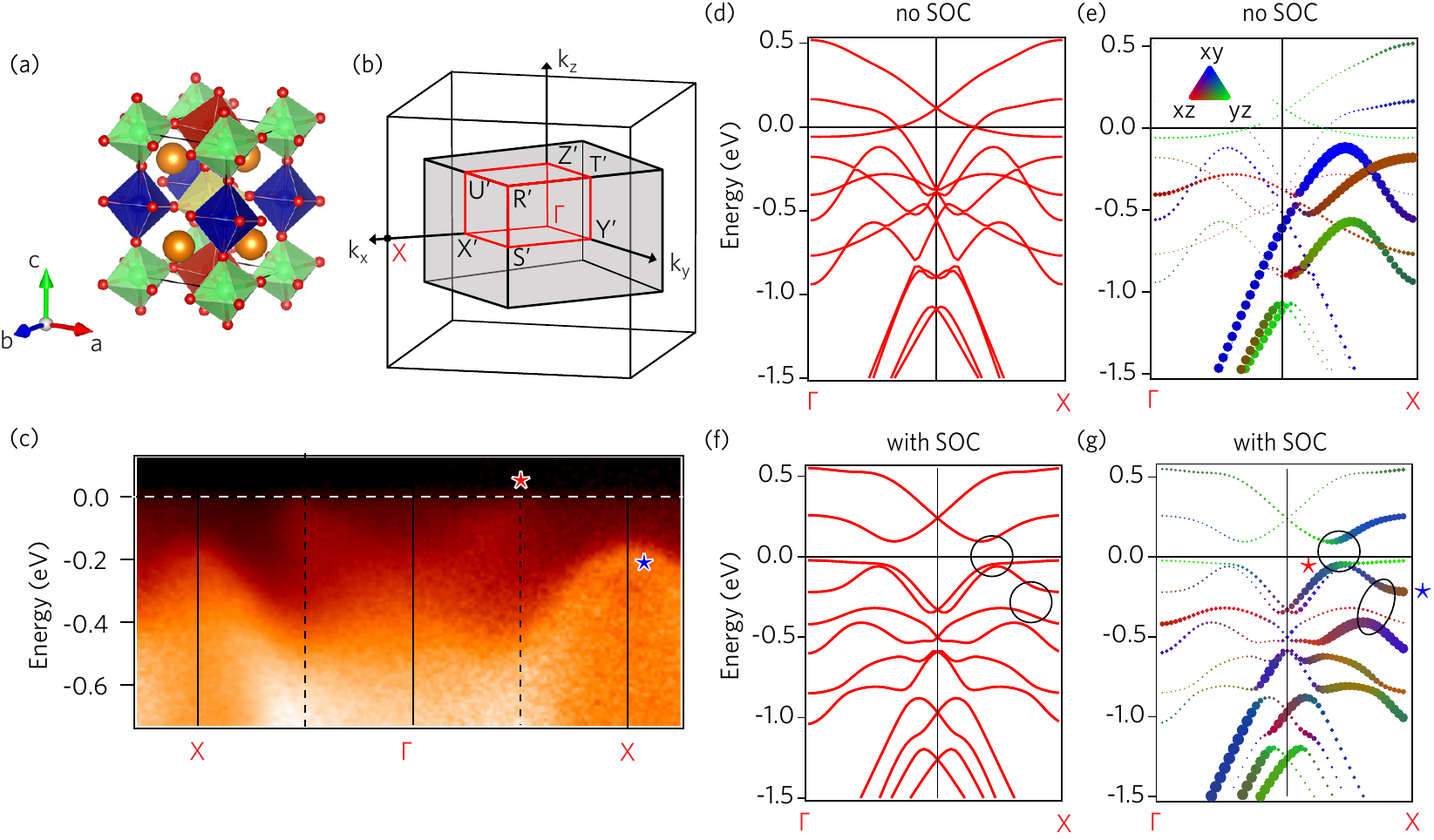}
    \caption{(a) Sketch of the \ce{SrIrO3} structure. There are 4 inequivalent Ir, at the center of oxygen octahedra of a different color. (b) Sketch of the corresponding BZ. The black cube corresponds to the primitive BZ (pseudocubic unit cell) and the shaded region to the supercell BZ. (c) Energy-momentum image of the dispersion along $\Gamma$X for a thin film of \ce{SrIrO3}/\ce{SrTiO3}, measured with 100~eV photon energy. The $\Gamma$X path is the diagonal of the primitive unit cell similar to Fig.~\ref{fig-BZ-Sr2IrO4}(c). (d) Calculated band structure along $\Gamma$X for $k_z=0$ without SOC. (e) Same as (d) with unfolding weight as marker size and color scale indicating $d_{xz}$ (red), $d_{yz}$ (green) and $d_{xy}$ (blue) character. (f,g) Same as (d,e) with SOC. The circles highlight the regions where a large SOC-induced gap opens. The stars highlight bands discussed in the paper.}\label{fig-SrIrO3}
\end{figure}

A very interesting case is the related 3D iridate \ce{SrIrO3}. Its structure is shown in Fig.~\ref{fig-SrIrO3}(a). In addition to in-plane rotation of the oxygen octahedra, similar to \ce{Sr2IrO4}, it also exhibits a tilt of the oxygen octahedra from the $c$ axis, inducing another type of folding along $k_z$. The resulting BZ is shown in Fig.~\ref{fig-SrIrO3}(b), it is rotated 45$^{\circ}$ in the $(k_x, k_y, 0)$ plane, as for \ce{Sr2IrO4}, but also halved along $k_z$. The calculated band structure is semimetallic and the four $J=1/2$ folded bands are expected to form a Dirac nodal line around the U point $(1/2,0,1/2)$ \cite{Carter_PRB_85_2012}. As topological features are rare in correlated oxides, this occurrence generated great interest. However, the orthorhombic structure of \ce{SrIrO3} is only stable in thin films, which adds questions on the role of the epitaxial constraint and the survival of topological features in real systems \cite{Liu_PRB_93_2016}.
It would therefore be very interesting to look for these features directly with ARPES, but the situation is not yet concluding. The only ARPES studies available to date were performed at a fixed photon energy \cite{Nie_PRL_114_2015,Liu_SR_6_2016}, which may not allow to precisely locate the U point, or at high photon energies, where the energy resolution is lower \cite{Schuetz_PRL_119_2017}. 

We only sketch here the help of unfolded calculations for deciphering the electronic structure of \ce{SrIrO3}, more details will be published later. Figure~\ref{fig-SrIrO3}(c) shows the images of the dispersion along $\Gamma$X direction with a photon energy of 100~eV obtained on a \ce{SrIrO3}(001) thin film ($t=10.4$~nm) grown on \ce{SrTiO3}(001) substrates (for clarity, we keep the same labelling as previously for the $(k_x, k_y, 0)$ plane of \ce{Sr2IrO4} in Fig. ~\ref{fig-BZ-Sr2IrO4}(c), i.e., X labels the corner of the primitive 2D BZ similar to Fig.~\ref{fig-BZ-Sr2IrO4}). The ARPES image looks quite simple, with a prominent band at X (almost invisible at $\Gamma$), resembling the $J=3/2$ in \ce{Sr2IrO4} shifted up by 0.2~eV and a band reaching the Fermi level at the middle of $\Gamma$X, looking like $J=1/2$ in \ce{Sr2IrO4} shifted up by 0.15~eV. Quantitatively, the comparison with raw calculations in the supercell is usually very confusing, as a much more complicated band structure is predicted [Fig.~\ref{fig-SrIrO3}(f)]. We will show how unfolding of the band structure can simplify the picture.  

We calculated with WIEN2k the electronic structure for the structure given in Table III of Ref.~\citenum{Puggioni_JAP_119_2016}. The space group is Pbnm (\#62), and the parameters are given in Table III. As for \ce{Sr2IrO4}, we can define the lattice vector matrix of the supercell ({\AA}) and the fictitious primitive cell, as well as the transformation matrix:
\begin{equation}
    A = 
    \begin{bmatrix}
        5.597 & 0 & 0 \\
        0 &  5.568 & 0 \\
        0 & 0 & 7.892
    \end{bmatrix},
    ~~a = 
    \begin{bmatrix}
        A_1/2 & -A_3/2 & 0 \\
        A_1/2 & A_2/2 & 0 \\
        0 & 0 & A_3/2
    \end{bmatrix},
    ~~P = 
    \begin{bmatrix}
        1 & -1 & 0 \\
        1 & 1 & 0 \\
        0 & 0 & 2
    \end{bmatrix}
\end{equation}
No Coulomb repulsion $U$ was used and a semi-metallic non-magnetic state is obtained, as it is the case experimentally \cite{Zhao_JAP_103_2008}.

Assuming our ARPES data at this photon energy correspond to $k_z=0$, we show the calculated electronic structure in Fig.~\ref{fig-SrIrO3}(d,f) with and without SOC. Clearly, the calculation looks much more complicated than ARPES data, and it is extremely difficult to understand which bands should be compared to the experiment. 

After unfolding [Fig.~\ref{fig-SrIrO3}(e,g)], a set of 3 bands is clearly emphasized, although their dispersions are affected by their mutual interactions. Using a color code for orbital characters (the \texttt{case.inq} file was modified to rotate $\mathbf{a}$ and $\mathbf{b}$ by 45$^{\circ}$ and align them with \ce{Ir-O-Ir} bonds), one can clearly recognize the original $d_{xz}$/$d_{yz}$/$d_{xy}$ bands [panel (e), without SOC] and $J=1/2$, $J=3/2$ after their interaction with SOC [panel (g)]. The places where a SOC induced hybridization gap openings are noted as empty circles. Obviously the band marked by a red star is at  significantly different position in the measurement, compared to the calculation. On the other hand, the band at X (blue star) exhibits a well-defined parabolic shape in the measurement over 0.4~eV, while there is a large SOC induced gap near the top of the dispersion in the calculation. Similar to \ce{Sr2IrO4}, we can anticipate that the effective spin-orbit coupling may be underestimated in the calculation, which will change the splitting of the two bands and also the position of their crossing, hence the break in the dispersion. Moreover, as the shape of $d_{xy}$ depends sensitively on the rotations, it may be necessary to tune the structure to get the right interaction pattern. This has extremely important consequences for the formation of the Dirac nodal line at $k_z=1/2$, which we will not discuss here. This examples shows how adjusting the structure to get a better description of the experimental data will be possible only when a basic understanding of the origin of the different bands is reached, thanks to the unfolding scheme.

Let us stress that the intensity of the folded bands is also a fine marker of the strength of the interactions at the origin of the supercell, which is here the rotations of the oxygen octahedra.  In a similar spirit, the information contained in the intensity of the folded bands on these interactions was recently used to refine the structure of \ce{SrRuO3} thin films \cite{Sohn_JKPS_s40042-022-00633-5_2022}. The near absence of the folded bands in our measurement suggests a small coupling. From the topological point of view, the meaning of the crossing of two bands with very different spectral weight is also a question that may deserve further work.

\section{Conclusion}

Unfolding of a supercell band structure into a primitive Brillouin zone is important for understanding implications of structural distortions, disorder, defects, solid solutions on materials electronic structure. Necessity of the band unfolding is also recognised in interpretation of angle-resolved photoemission spectroscopy (ARPES) measurements. We described an extension of the \texttt{fold2Bloch} package by implementing an arbitrary transformation matrix used to establish a relation between primitive cell and supercell. The convention selected for the transformation matrix is compatible with that recommended by the International tables for crystallography. This development allows us to overcome limitations of supercells constructed exclusively by scaling of primitive cell lattice vectors. For instance, it becomes possible to transform between primitive and conventional cells as well as include rotations. The updated \texttt{fold2Bloch} package is available from a GitHub repository as a FORTRAN code. It interfaces with the all-electron full-potential WIEN2k and the pseudopotential VASP density functional theory packages. The \texttt{fold2Bloch} is supplemented by additional pre- and post-processing utilities that aid in generating k points in the supercell (such that they later fall onto a desired path in the primitive Brillouin zone after unfolding) and plotting the unfolded band structure. We selected \ce{Sr2IrO4} as an illustrative example and, for the first time, present its properly unfolded band structure in direct comparison with ARPES measurements. In addition, critical importance of the band unfolding for interpretation of \ce{SrIrO3} ARPES data is illustrated and discussed as a perspective. Of particular interest for iridates is the ability of ARPES to sense imprints that tilting of \ce{IrO6} octahedra leaves on the materials' electronic structure. ARPES measurements teach us an important lesson: small structural perturbations neither lead to a sudden change in the electronic structure nor redefine the associated primitive Brillouin zone, contrary to what is expected from the formal symmetry of the structure.

\begin{acknowledgments}
Calculations were performed using the Compute Canada infrastructure supported by the Canada Foundation for Innovation under John R. Evans Leaders Fund. ARPES work was supported by the Agence Nationale de la Recherche grant ``SOCRATE" (Grant No. ANR-15-CE30-0009-01).
\end{acknowledgments}


%

\clearpage
\section*{Supporting information}

We include a ZIP archive with WIEN2k structures, relevant scripts, initialization and unfolding workflows. See README file within the archive for additional description of the content.


\clearpage

\clearpage

\clearpage

\clearpage

\clearpage

\clearpage

\clearpage

\clearpage

\begin{table}[h]
\caption{Structural and calculation parameters.}
\label{tab-param}
\begin{ruledtabular}
\begin{tabular}{ l c c}
Parameters	& \ce{Sr2IrO4} & \ce{SrIrO3}\\
\colrule
Space group (non-magnetic) & 88 (I4$_1$/a) & 62 (Pbnm) \\
Space group (magnetic) & 13 (P2/c) & n/a \\
Lattice param. ({\AA}) & 5.485, 25.775 & 5.568, 5.597, 7.892 \\
$R_\text{MT}$ (bohr) & 2.24 (Sr) & 2.26 (Sr)\\
& 1.98 (Ir) & 2.09 (Ir) \\
& 1.62 (O) & 1.71 (O)\\
$n_\text{val}$ & 10 (Sr) & 10 (Sr) \\
& 31 (Ir) & 31 (Ir) \\
& 2 (O) & 2 (O) \\
$R_{\text{MT}_{\text{min}}}K_{\text{max}}$ & 7.0 & 7.0 \\
$G_{\text{max}}$ & 12 & 12 \\
$l_{\text{max}}$ & 10 & 10  \\
$l_{\text{vns}_{\text{max}}}$ & 6 & 4 \\
$k$ mesh & $12\times 12 \times 2$ & $11\times 11 \times 7$  \\
& (shifted) \\
Energy (Ry) and & $10^{-4}$ & $10^{-4}$  \\
charge converg. & $10^{-3}$ & $10^{-3}$ \\
\end{tabular}
\end{ruledtabular}
\end{table}

\clearpage

\begin{table}[h]
\caption{BCT reciprocal lattice points.}
\label{tab-BCT-k-conversion}
\begin{ruledtabular}
\begin{tabular}{ l c c c}
Label & Conventional \cite{Aroyo_ACA_70_2014} & Primitive & Primitive \\
&& (standard \cite{Aroyo_ACA_70_2014}) & (Fig.~\ref{fig-BZ-Sr2IrO4}c) \\
\colrule
$\Gamma$ & (0, 0, 0) & (0, 0, 0) & (0, 0, 0) \\
X & (1/2, 1/2, 0) & (0, 0, 1/2) & (1/2, 1/2, 1/2)  \\
M & (1/2, 0, 0) & ($-$1/4, 1/4, 1/4) & (1/2, 0, 1/4)  \\
\end{tabular}
\end{ruledtabular}
\end{table}

\end{document}